\documentclass[journal = jpclcd]{achemso}
\usepackage{graphics}
\usepackage{amsmath}
\usepackage{braket}
\usepackage{graphicx}
\usepackage{enumerate}
\usepackage{xcolor}
\usepackage{amssymb}
\usepackage{xr}
\makeatletter

\newcommand*{\addFileDependency}[1]{
  \typeout{(#1)}
  \@addtofilelist{#1}
  \IfFileExists{#1}{}{\typeout{No file #1.}}
}
\makeatother

\newcommand*{\myexternaldocument}[1]{
    \externaldocument{#1}
    \addFileDependency{#1.tex}
    \addFileDependency{#1.aux}
}

\myexternaldocument{SI}
\definecolor{LightCyan}{rgb}{0.88,1,1}
\definecolor{LightGreen}{rgb}{0.26, 0.97, 0.60}
\definecolor{LightRed}{rgb}{1, 0.47, 0.36}

\usepackage[normalem]{ulem}

\author{Carlos~Mejuto-Zaera}
\affiliation{University of California, Berkeley, California 94720, United States}
\author{Guorong~Weng}
\affiliation{University of California, Santa Barbara, California 93106, United States}
\author{Mariya~Romanova}
\affiliation{University of California, Santa Barbara, California 93106, United States}
\author{Stephen~J.~Cotton}
\affiliation{Quantum Artificial Intelligence Lab. (QuAIL), Exploration Technology Directorate,
NASA Ames Research Center, Moffett Field, CA 94035, USA}
\affiliation{KBR, 601 Jefferson St., Houston, TX 77002}
\author{K.~Birgitta~Whaley}
\affiliation{University of California, Berkeley, California 94720, United States}
\author{Norm~M.~Tubman}
\affiliation{Quantum Artificial Intelligence Lab. (QuAIL), Exploration Technology Directorate,
NASA Ames Research Center, Moffett Field, CA 94035, USA}
\author{Vojt\v{e}ch~Vl\v{c}ek}
\affiliation{University of California, Santa Barbara, California 93106, United States}
\email{vlcek@ucsb.edu}

\title{Are multi-quasiparticle interactions important in molecular ionization?}


\begin{document}

\makeatletter
\setlength\acs@tocentry@height{4.4cm}
\setlength\acs@tocentry@width{12.1cm}
\makeatother

\begin{tocentry}
    \includegraphics[scale=0.5]{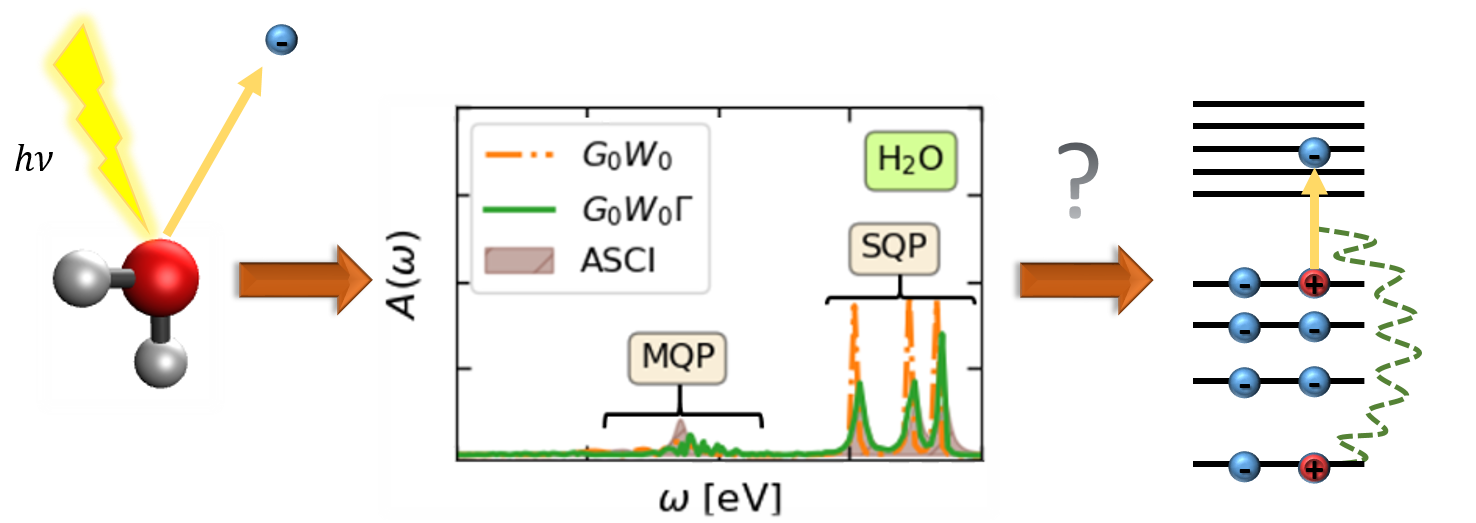}
\end{tocentry}

\date{\today}

\maketitle

\begin{abstract}
Photo-emission spectroscopy directly probes individual electronic states, ranging from single excitations to high-energy satellites, which simultaneously represent multiple quasiparticles (QPs) and encode information about electronic correlation. First-principles description of the spectra requires an efficient and accurate treatment of all many-body effects. This is especially challenging for inner valence excitations where the single QP picture breaks down. Here, we provide the full valence spectra of small closed-shell molecules, exploring the independent and interacting quasiparticle regimes, computed with the fully-correlated adaptive sampling configuration interaction (ASCI) method.  We critically compare these results to calculations with the many-body perturbation theory, based on the $GW$ and vertex corrected $GW\Gamma$ approaches. The latter explicitly accounts for two-QP quantum interactions, which have been often neglected. We demonstrate that for molecular systems, the vertex correction universally improves the theoretical spectra, and it is crucial for accurate prediction of QPs as well as capturing the rich satellite structures of high-energy excitations. $GW\Gamma$ offers a unified description across all relevant energy scales.  Our results suggest that the multi-QP regime corresponds to dynamical correlations, which can be described via perturbation theory.  
\end{abstract}

\maketitle
The quantitative understanding of electronic excitations in complex molecular extended systems is one of the most fundamental open challenges in modern theoretical chemistry. Perhaps the most direct experimental probe of the excited state manifold is given by photo-emission spectra (PES), which directly access individual electronic states~\cite{Hufner1984,Sander1987,Chewter1987,Wu2011} and, in principle, access information on the single-particle orbitals~\cite{Puschnig2009,Luftner2014,Vozzi2011}. Theoretical approaches are meant to provide an interpretative connection between measured spectral features and chemical concepts, thus helping design systems with tailored (opto)electronic properties. An accurate realization of this ideal often requires a treatment beyond effective one-body theories, such as Hartree-Fock or Kohn-Sham density-funcitonal theory,  to capture all interactions that renormalize single electron properties and lead to the formation of quasiparticles (QPs). 

Traditional gold-standard wave function approaches, such as the configuration interaction or coupled cluster methods, can yield highly accurate predictions, and they have been extensively applied to describe the outer valence and core electron spectra of small molecules~\cite{Bagus1977,Lisini1991,Honjou1981,Peng2018,Nishi2018,Rehr2020}, which fundamentally behave as individual QPs. However, the description of the inner valence region is more complicated: multiple excitation mechanisms are available, and ionization can be accompanied by simultaneous neutral excitation of the system. At the energy scales associated with inner valence PES, the single QP (SQP) picture breaks down, and the spectra exhibit a multitude of features~\cite{Cederbaum1977,Cederbaum1980}, which we refer to as the multi-QP (MQP) regime.  In analyzing PES, these are typically referred to as shake-up satellites. Describing these ionized states requires accounting for the nontrivial interactions among many excited states. The inner valence excitations in the MQP regime thus represents an unambiguous measure of electron correlation~\cite{Zhou2020}. Ab-initio wave function methods can capture the high energy holes in principle, but they often require extensive exploration of the exponentially large Hilbert space, limiting their application to small molecular systems. Further, while convenient from a computational perspective, the formulation of these methods in terms of determinants makes a definite distinction between dynamic and static correlation complicated in most cases. In particular, it is unclear whether MQP features in PES belong to the former or latter class. On the one hand, they are intrinsically many-body effects which cannot be captured by a single Slater determinant. On the other hand, they often arise from neutral excitations weakly coupled to holes, which suggests they should be amenable to perturbative description.

In the pursuit of an accurate but scalable theoretical description of PES, many-body perturbation theory (MBPT) has proven a viable alternative for computing QP excitations. It offers a systematically improvable theoretical framework for capturing dynamical correlations. MBPT relies on physically motivated concepts such as screening, while retaining a polynomial computational scaling. In particular, the popular $GW$ approximation is extensively applied to treat molecules and solids, and it predicts outer valence and core electron spectra with good accuracy~\cite{Golze2019,Bruneval2013,Blase2011,Vlcek2017,Van2015,Golze2018}. With recent algorithmic developments, $GW$ can be applied to systems with thousands of electrons~\cite{Neuhauser2014,Vlcek2017,Vlcek2018,Vlcek2018-PRM,Romanova2020,Brooks2020,Weng2020}. Unfortunately, the $GW$ framework is fundamentally limited to systems where classical electrodynamics dominates the electronic correlation. Hence, the description of the MQP regime is beyond the $GW$ capabilities~\cite{Romaniello2012, Maggio2017,Vlcek2019,martin_reining_ceperley_2016,Onida2002,Marini2004,Bruneval2005,Schindlmayr1998}. Within the MBPT framework, the missing higher-order quantum interactions are represented by the vertex term $\Gamma$, which encodes dynamical two-particle correlations. The inclusion of $\Gamma$ leads to the $GW\Gamma$ approach, which for molecular systems has been studied only on first ionization energies and electron affinities\cite{ delSole1994,Shishkin2007, Hellgren2018, Springer1998, Maggio2017,Vlcek2019}. However, the role of $\Gamma$ on the entire valence states and the MQP excitations has not been explored up to now. 

In this letter, we demonstrate that MBPT with vertex terms successfully captures the correlations necessary for a qualitative the description of the MQP regime of PES, suggesting them to be dynamic in nature. By comparison with explicitly correlated and numerically exact methods, namely the adaptive sampling configuration interaction (ASCI) approach~\cite{Tubman2016,Tubman2018a,Mejuto2019,Tubman2020}, we show that the  $GW\Gamma$ approach yields superior results compared to $GW$ throughout all energy scales. Moreover, we discover that the non-local quantum interactions play a significant role for all valence electron excitations and correctly captures the breakdown of the SQP picture, eliminating spurious artifacts of $GW$  in the shake-up region of PES. We choose as case study the PES of selected closed shell molecules ($\mathrm{NH_3}$, $\mathrm{H_2O}$, $\mathrm{CH_4}$, $\mathrm{C_2H_2}$, and $\mathrm{N_2}$), which as we show present rich MQP character in the inner valence region.

The  spectral function $A(\omega)$ computed with ASCI and MBPT is the figure of merit. This corresponds to the physical observable (i.e., PES ), and is given as the trace over the imaginary part of the Green's function (GF) $G_{i,j}(\omega)$~\cite{martin_reining_ceperley_2016}, i.e., ${A(\omega) = -\frac{1}{\pi}\mathrm{tr}\left[\operatorname{Im} G(\omega)\right]}$. The $i,j$ sub-indices correspond to a chosen single particle basis. The peaks in $A(\omega)$ correspond to the poles of $G$~\cite{Fetter2003} and for a particular hole component:
\begin{equation}
    G_{i,i}(\omega) = \sum_m \frac{\left|\braket{\Psi_m^{N-1}|c_i|\Psi_0^N}\right|^2}{\omega + \left(E_m^{N-1} - E_0^N\right) -i\eta},
\label{eq:GF}
\end{equation}
where $\ket{\Psi_{m}^{N}}$ is the $m$-th eigenstate of the $N$ particle system, with energy $E_m^N$. Further, $c_i$ is the $i$-th annihilation operator and $\eta$ is an integrating factor. 
In the mean-field and SQP regimes, we expect only \emph{one}   non-vanishing overlap ${\left|\braket{\Psi_m^{N-1}|c_i|\Psi_0^N}\right|^2}$ per single-particle state $i$, corresponding to a dressed hole with energy $E_m^{N-1} - E_0^N$. However,  \emph{multiple} states $\ket{\Psi_m^{N-1}}$ with distinct  $E_m^{N-1}$ exist in the MQP regime. 

First, we evaluate eq~\ref{eq:GF} without further approximations via the ASCI algorithm~\cite{Tubman2016,Tubman2018a,Tubman2019,Mejuto2019,Tubman2020}, which captures both SQP and MQP regimes, regardless of the degree of correlation of the excited states. ASCI provides accurate Green's functions in model systems~\cite{Mejuto2019,Mejuto2020}; here, we compute Green's functions of \emph{ab initio} Hamiltonians using ASCI for the first time. The spectral features calculated from eq~\ref{eq:GF} represent a series of infinitely sharp peaks due to the use of finite atomic basis sets and a Hermitian Hamiltonian. As a result, no scattering states are considered and the finite lifetimes  of individual excitations (arising due to the coupling to continuum) are neglected. To facilitate comparison with the MBPT results, which do include lifetimes due to electron-electron scattering, the ASCI-computed spectra have been artificially broadened using the $\eta$ parameter in eq~\ref{eq:GF}. For the heuristics and technical details of ASCI, including basis set extrapolation, see the Supporting Information (SI)~\cite{supp}.

MBPT is based on the conceptually different approach of obtaining the QP energies through the Dyson equation~\cite{martin_reining_ceperley_2016}.  This relates the Green's function of the fully interacting QPs $G_{i,j}(\omega)$ and a reference (mean-field) Green's function $G_{0,i,j}(\omega)$, through the self-energy $\Sigma_{i,j}(\omega)$. The latter, in principle, contains all many-body effects. The poles of $G_{i,j}(\omega)$ are subject to the QP fixed point equation:
\begin{equation}
    \epsilon_j = \epsilon_j^0 +  \operatorname{Re}\left[\Sigma_{j,j}(\omega = \epsilon_j)\right] + \operatorname{Re}\left[\Delta_{j}(\omega = \epsilon_j)\right].
\label{eq:QPE}
\end{equation}
Here $\epsilon_j^0$ is a pole of $G_0$ and $\Delta_{j}(\omega)$ comprises the coupling due to the off-diagonal elements of $\Sigma_{i,j}(\omega)$. In the SQP regime, $\Sigma(\omega)_{i,j}$ merely shifts the poles of $G$ with respect to $G_0$, and eq.~\ref{eq:QPE} only has one solution per orbital. For MQPs, the structure of the self-energy is necessarily more complex and, in principle, yields multiple solutions to eq~\ref{eq:QPE}. To capture such correlated states, $\Sigma(\omega)_{i,j}$ must account for the interactions between the particle-hole pairs and ionized holes (or injected electrons) in the system. It is often helpful to represent the solutions of eq.~\ref{eq:QPE} graphically as done in the discussion below. 

In this work, we base the perturbation expansion on top of the Hartree-Fock (HF) Hamiltonian. The correlation self-energy, $\Sigma_c$, is derived from the equation of motion of the QP Green's function and the non-interacting reference (i.e., HF in this case). The full expression reads~\cite{martin_reining_ceperley_2016}:
\begin{equation}
\Sigma_c(1,2) = -\nu(1,\bar4)G(1,\bar3)\frac{\delta\Sigma_{Hxc}(\bar3,2)}{\delta{G(\bar6,\bar5)}}{^3\chi(\bar6,\bar5,\bar4)},    
\label{correlation2}
\end{equation}
where we employ a short-hand notation for space-time coordinates $1\equiv (r_1, t_1)$ and bar indicates a coordinate that is integrated over. Here, $\Sigma_{Hxc}$ contains Hartree, exchange, and correlation interactions, and $\nu(1,2) = \delta(t_1-t_2)/|r_1 - r_2|$ is the bare Coulomb interaction. Further, we introduce a generalizes response function connecting the induced Green's function to a perturbing potential $U(1)$ as ${^3\chi(1,2,3) = -i\delta G(1,2)/ \delta U(3)}$.\cite{Vlcek2019} To lower the computational cost, it is common to approximate ${\delta\Sigma_{Hxc}}/{\delta{G}}$, the two-particle interaction kernels~\cite{Onida2002,martin_reining_ceperley_2016,Marini2004,Bruneval2005,Schindlmayr1998}. In particular, taking ${\delta\Sigma_{Hxc}}/{\delta{G}} \approx {\delta\Sigma_{H}}/{\delta{G}}$ leads to the popular $GW$ approximation~\cite{Hedin1965,Aryasetiawan1998}, which describes the correlations as induced time-dependent density fluctuations. In contrast, $GW\Gamma$ evaluates eq.~\ref{correlation2} in full and captures the QP couplings. The additional terms (compared to $GW$) stemming from ${\delta\Sigma_{xc}}/{\delta{G}}$ are referred to as vertex corrections, and they are responsible for mutual coupling of MQP interactions~\cite{Romaniello2012, Maggio2017, Vlcek2019,martin_reining_ceperley_2016,Bruneval2005}. In particular, it has been shown that including vertex corrections is necessary for self-consistent renormalization\cite{Shirley1996} and to capture satellite spectral features for simple models of core electrons~\cite{Ness2011}. While the computational cost of $\Gamma$ is large~\cite{Grueneis2014,Maggio2017}, a recent stochastic formalism introduced a linear-scaling algorithm, which we apply here~\cite{Vlcek2019} to study the effect of vertex corrections on the satellite features in the valence spectrum. 

In practice, we resort to the one-shot correction scheme for both $GW$ and $GW\Gamma$. In this approximation the two-particle kernel derives from the mean-field Hamiltonian, i.e., it becomes ${\delta\Sigma_{H}}/{\delta{G}}$ for  $GW$. For $GW\Gamma$, the kernel is ${\delta\Sigma_{Hx}}/{\delta{G}}$, i.e.,  it also includes the HF exchange term~\cite{Vlcek2019}. Further, as common, we employ the random-phase approximation (RPA) in $GW$, while we avoid it in $GW\Gamma$, thus accounting for excitonic effects in the response function. The use of $GW$ without RPA is possible, but it worsens the quality of $A(\omega)$~\cite{Vlcek2019}.  Unlike other starting points, HF is not associated with spurious multiple solutions to eq.~\ref{correlation2}~\cite{Golze2020}. Further, HF single-particle states are close to true Dyson orbitals~\cite{Diaz-Tinoco2019}. Hence, HF provides an ideal mean-field reference for exploring the capabilities of MBPT.

\begin{figure}
\centering
\includegraphics[width=\linewidth]{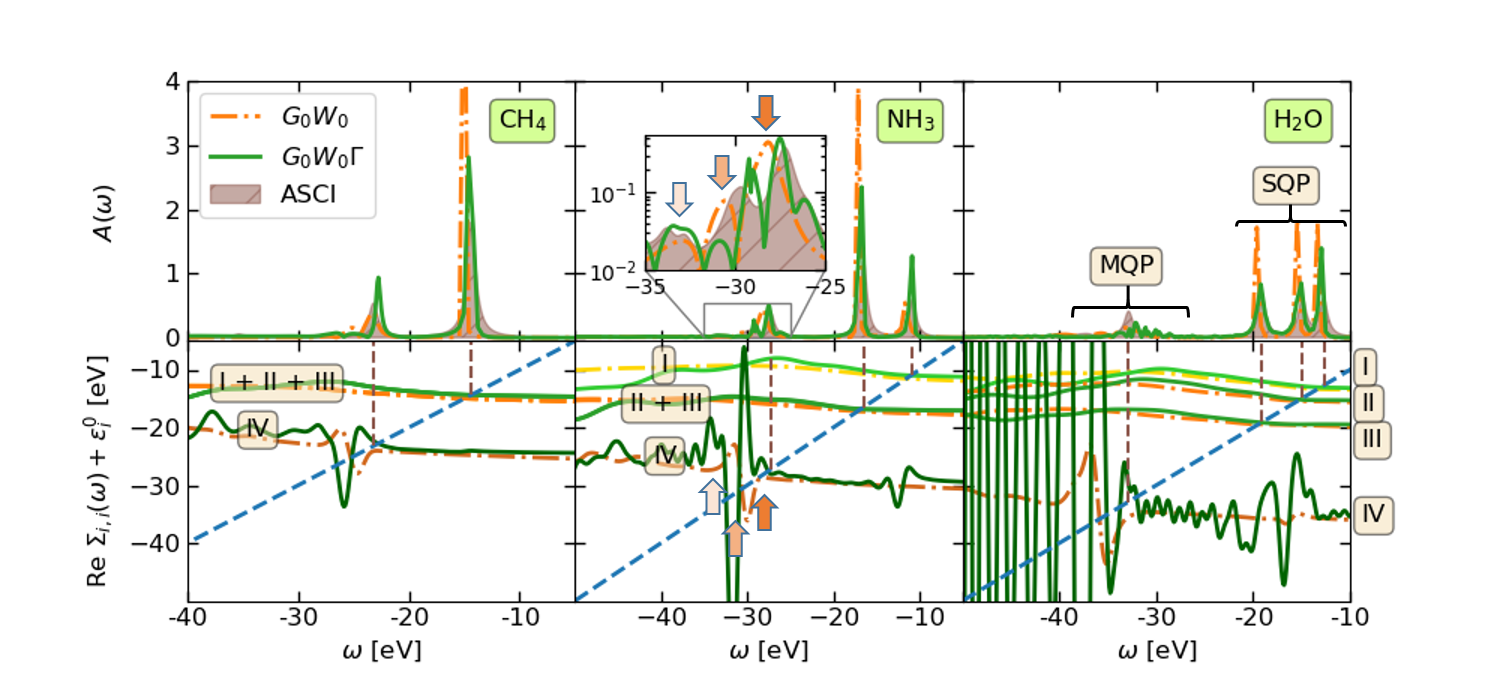}
\caption{\label{fig:MBPTvsASCI_IV} Upper panels: Spectral functions for CH$_4$, NH$_3$ and H$_2$O as computed with ASCI (filled curve), $G_0W_0$ (dot-dashed line) and $G_0W_0\Gamma$ (solid line). Note that the inset in the central panel is in log scale. We mark the SQP and MQP regimes explicitly for H$_2$O. Lower panels: Corresponding real part of the diagonal self-energy terms from $G_0W_0$ and $G_0W_0\Gamma$, one curve per orbital numbered starting with I for the HOMO, shifted by corresponding Hartree-Fock QP energy $\epsilon_i^0$. Symmetry induced degeneracies reduce the number of peaks and self-energy curves in CH$_4$ and NH$_3$. The former has a three-fold degenerate HOMO at $\sim-14$~eV, and the latter a two-fold degenerate at $\sim-17$~eV. The vertical dashed lines mark the position of the ASCI QP energies, and the blue dashed line corresponds to the frequency line $y(\omega) = \omega$. The intercepts of the self-energy curves with the blue dashed line correspond to graphical solutions of the QP equation~\ref{eq:QPE}. The arrows point at the features in the $G_0W_0$ highest energy curve which create maxima in the spectral function, see text for discussion.}
\end{figure}

The computed spectral functions with the different theoretical approaches are illustrated in the top panel of Figure~\ref{fig:MBPTvsASCI_IV} for the CH$_4$, NH$_3$, and H$_2$O molecules. The frontier orbitals appear at the lowest absolute frequencies. The distinction between the SQP and MQP regimes is evident: the highest occupied molecular orbitals (HOMO) and states energetically close to these are composed of a single sharp peak per orbital, consistent with the SQP picture. In contrast, excitations far away from the HOMO exhibit broader peaks; their spectral intensity is often redistributed to multiple satellite features. This is a signature of the MQP regime. For the first time, we are able to study this behavior in molecules with a fully correlated approach. 
These results alone however do not indicate whether the satellites represent a weak coupling of, in principle, distinguishable QPs, and to what extent the MQP couplings can be captured perturbatively using single particle states forming only one Slater determininant.

\begin{figure}
\centering
\includegraphics[width=\linewidth]{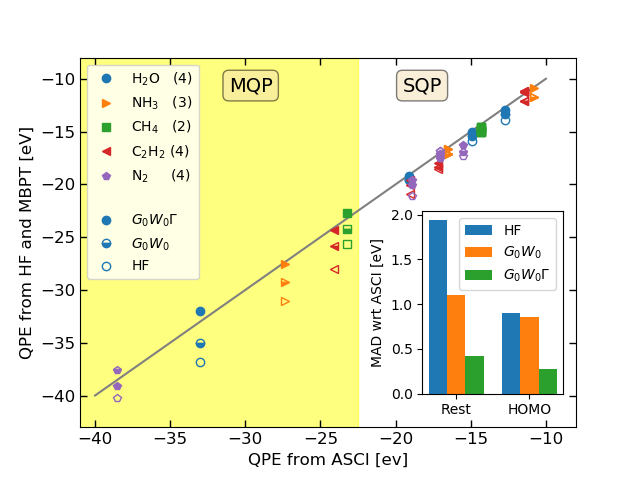}
\caption{\label{fig:ASCIvsMBPT} Quasiparticle energies (QPE) for the different molecular systems computed from ASCI vs the QPEs computed with the many-body pertubation theory. The symbols correspond to: Hartree-Fock (HF, empty), $G_0W_0$ (half-filled), and the vertex corrected $G_0W_0\Gamma$ (solid). The energy region where the excitations present MQP character is shaded yellow. The number of QPEs per molecule is given in the legend. The inset shows the mean average deviation (MAD) between the ASCI results and the three perturbative approaches.}
\end{figure}

To answer this, we compare the ASCI and MBPT calculations.
We show the QP energies for all valence excitations and all methods in Figure~\ref{fig:ASCIvsMBPT}. As expected~\cite{Onida2002,Grueneis2014,Maggio2017,Vlcek2019,Marini2004,Bruneval2005, Schindlmayr1998}, $GW\Gamma$ is closest to ASCI results (compared to HF and $GW$) and we find the best agreement for the holes of the HOMO states, which are in the SQP regime. For HOMO, the self-energy is merely responsible for shifting the poles of the Green's function, but the presence of the vertex corrections is important, as  illustrated in the inset of Figure~\ref{fig:ASCIvsMBPT}. The one-shot $GW$ approach performs only slightly better than HF; the mean absolute deviation (MAD) with respect to ASCI is $1.0$~eV  and $0.8$~eV for HF and $GW$, respectively. Upon inclusion of vertex terms, however, the MAD decreases to $0.3$~eV.  The presence of $\Gamma$ is responsible for the excitonic effects in $W$, and these are likely non-negligible in small systems, where electrons and holes have large spatial overlaps. Further, the vertex correction cancels, at least partially, the spurious self-polarization in $GW$~\cite{Nelson2007, Aryasetiawan2012}. The latter is possibly the major driving force of the improvement, because the electron-hole interactions in the screening tend to have little or (surprisingly) negative impact on the QP energy predictions~\cite{Maggio2017, Lewis2019, Vlcek2019}.

Unlike previous studies, we go beyond the HOMO and also examine higher energy excitations, which exhibit MQP behavior. Due to the lack of correlation, the HF ionization potentials deviate significantly from the ASCI QP energies (MAD $\sim 2$~eV), as illustrated in Figure~\ref{fig:ASCIvsMBPT}. Even here, the ASCI spectral function retains a dominant peak (in Figure~\ref{fig:MBPTvsASCI_IV} at -23~eV for CH$_4$, -27~eV for NH$_3$ and -33~eV for H$_2$O); yet, a fraction of the spectral weight is redistributed to satellite features. This departure from the SQP regime progresses with the increasing excitation energy. For the highest energy valence states of  H$_2$O and N$_2$ (at -33~eV and -39~eV respectively),  $A(\omega)$ shows multitude of sizeable satellites located $>10$~eV away from the main peak. Unsurprisingly,  the MBPT QPEs deviate more strongly from ASCI in this regime, see inset of Figure~\ref{fig:ASCIvsMBPT}. Nevertheless, for these higher energy states the vertex correction again significantly improves the QPEs of the most prominent peaks, bringing the MAD below 0.5 eV. Considering the main QP signatures alone, the vertex corrections seem necessary for an accurate description throughout all studied energy scales, suggesting that MQP effects are important even for the simple orbital ionization energies.

We now turn to address the satellites in the MQP regime. Processes at this energy scale are characterized by the presence of neutral (e.g. optical) excitations interacting with the ionized hole. The distinction between weakly and strongly interacting MQPs is not possible by inspecting $A(\omega)$ alone, since it does not affect the shape of the peaks in a characteristic way. In the rest of this paper, we explore how the  MBPT methods may offer some clarification by investigating the properties of $\Sigma(\omega)$, since the poles of $A(\omega)$ are related to $\Sigma(\omega)$ through eq~\ref{eq:QPE}.

A graphical solution to eq~\ref{eq:QPE} is illustrated in the lower panels of Figure~\ref{fig:MBPTvsASCI_IV}. The $GW$ $A(\omega)$ in the MQP regime shows a common simple structure: besides the main peak, there are two other smaller peaks accompanying it (see the arrows in the inset of Figure~\ref{fig:MBPTvsASCI_IV}). This three-peak feature corresponds to the self-energy with only a single true pole (middle arrow in lower central panel of Figure~\ref{fig:MBPTvsASCI_IV}), but having two nearby frequencies where eq~\ref{eq:QPE} is approximately fulfilled (left and right arrows). In almost all of the molecules, the main maximum corresponds one of these two ``pseudo-poles'' (rightmost arrow). This same three-peak structure has been discussed in the context of solids as an arifact of $GW$.~\cite{Aryasetiawan1996,Lischner2013, Lischner2015}. Note that due to the finite real-time propagation employed in our stochastic implementation~\cite{Neuhauser2014, Vlcek2017,Vlcek2018-PRM,Vlcek2018, Vlcek2019}, some of the $GW$ ``pseudo-poles'' may correspond to actual poles. Only for N$_2$, does eq~\ref{eq:QPE} have two solutions (i.e., two poles). Here the resulting $A(\omega)$ features only a single prominent peak associated with the low-energy QP~\cite{supp}. 
To summarize, ASCI shows a complex satellite structure, whereas $GW$  only presents a regular three-peak spectra for the bottom valence states.

Clearly, $GW$ can provide multiple solutions, in principle, though they do not match the fully correlated results.\cite{Guzzo2011,martin_reining_ceperley_2016,Lischner2015} The QPs predicted by $GW$ in closed shell systems is consistent with a \emph{surmised} ``plasmaron'', representing a resonantly bound hole and a collective neutral excitation. In solids, this was interpreted as an electron-plasmon state. However this was eventually identified as an artifact of $GW$~\cite{Langreth1970}. In practice, the $GW$ approximation spuriously substitutes the multiple satellites with a single secondary QP that does not correspond to a physical excited state. Nevertheless, for \emph{weakly} interacting MQPs, the absence of satellites can be remedied by reconstructing the Green's function  (and $A(\omega)$) via the cumulant expansion technique~\cite{Vlcek2018-PRM,Lischner2015}. This method reproduces MQP structures, but it inherently assumes the presence of a \emph{distinguishable} neutral excitation, which corresponds to a pole in the (classical) polarizability, which describes the charge density fluctuations. This in turn implies that $\ket{\Psi_m^{N-1}}$ in eq~\ref{eq:GF} has a direct product state of an ionized hole and excited state determinant; the total energy of such a state is merely the sum of the ionization potential and the neutral excitation energy. If true, the $A(\omega)$ would exhibit a regular satellite structure, in which peaks appear at energies corresponding to the multiples of the ``plasmon'' energy. In other words, the $n$-th satellite maximum represents the energy of a single hole plus $n$ excited plasmons. The weak coupling regime is justified for localized holes in the presence of delocalized neutral excitations~\cite{Langreth1970}. However, this separation is hard to conceive in small finite systems and cannot be  readily justified for molecules. Furthermore, the fully correlated ASCI results do \emph{not} exhibit a regular pattern of satellite peaks, suggesting that the MQP regime comprises at least some entangled $\ket{\Psi_m^{N-1}}$ states.

The two particle interactions present in $GW\Gamma$ enable mutual couplings between holes and neutral excitations. Hence, if the MQP regime stems from the dynamical correlation, the frequency dependence of the self-energy should capture it.  Indeed, the $GW\Gamma$ yields \emph{multiple peaks} in $A(\omega)$, recovering the same kind of rich satellite structure present in the ASCI spectra. To understand why, we illustrate $GW\Gamma$ self-energy in Figure~\ref{fig:MBPTvsASCI_IV} for CH$_4$, NH$_3$ and H$_2$O. By inspecting the results closely, we see that the distant satellites appear either: (i) due to the small denominator in eq~\ref{eq:GF}.\footnote{The distant peak stem from a combination of the small denominator in eq~\ref{eq:GF} in combination with the small imaginary part of $\Sigma$, which is not shown in Figure~\ref{fig:MBPTvsASCI_IV}. Further note that  the finite real-time propagation leads to numerical broadening; as a result, some of the features in the self-energy may not appear in the graphical solution as true poles, but only ``pseudopoles''.}, or (ii) because eq~\ref{eq:QPE} is satisfied at multiple frequencies.

Regardless of the system, the vertex correction is responsible for new features in $\Sigma$, especially in the inner-most valence energy regions.  For NH$_3$, the double peak in the $\sim -30$ eV region is accompanied by small satellites at lower energies stemming from the rapid variation of $\Sigma$.  The oscillatory behavior increases  with the energy of the excitated holes. This is clearly seen for the bottom valence region in H$_2$O and N$_2$: their inner valence states are at lower energies than NH$_3$ and the self-energy indeed exhibits stronger oscillations leading to a plethora of poles (more than ten) with energies almost 30~eV below the main QP peak (see Figure~\ref{fig:MBPTvsASCI_IV} and the SI).\footnote{ Note that the effect of the vertex is possibly overestimated due to the absence of self-consistency. The self-consistent solution  contains screened Coulomb interactions in $\Gamma$~\cite{Schindlmayr1998,Onida2002}. This renormalization will weaken the MQP couplings, reducing the oscillatory behavior shown in H$_2$O and N$_2$, retaining the envelope of their $A(\omega)$ and resulting in a smooth behavior, like for the other molecules.}
Further, we observe that the vertex corrections introduce ``pseudo-poles'' in the self-energy of the highest energy hole states that are resonant with the QP peaks of the lower energy holes; c.f. the ``pseudo-pole'' at -11~eV in the highest energy hole state (IV) of NH$_3$, resonant with the HOMO (I) QPE. This seems to underline the inter-orbital couplings arising from the vertex correction, but may be related to a mixed orbital character, see Supplementary Information (SI)~\cite{supp}.

In this letter, we theoretically investigated the electronic correlation which affects valence ionization energies and is directly linked to photoemission experiments. We provided virtually-exact ab-initio valence spectra for small molecular systems and explored whether and how many-body perturbation theory captures the various excitations. Namely, we compared the spectral functions computed with one-shot perturbative corrections obtained with the $GW$ and vertex-corrected $GW\Gamma$ approaches to adaptive-sampling CI. For the first time, we provide such a comparison for the entire range of valence excitations, i.e., the outer and inner valence holes and shake-up satellites.

We show that the neglect of explicit two-particle interactions in $GW$ leads to substantial errors. In contrast, $GW\Gamma$ results are close to the fully correlated calculations and exhibit a rich structure of multi-quasiparticle excitations stemming from the coupling between holes and optically excited states. While $GW$ yields spurious solutions, the presence of vertex terms removes the artifacts and correctly reproduces the peak structure in $A(\omega)$. 

The high-energy regime is typically associated with the breakdown of the quasiparticle picture. However, our results clearly show that the shake-up satellite features arise due to dynamical correlations. In other words, they can be described perturbatively using single-particle states of only one Slater determinant. Our findings should encourage the further development of perturbative methods that explicitly account for mutual multi-quasiparticle excitations via vertex terms. Beyond small molecular systems, this will have a decisive effect on the path towards a first-principles understanding of excited states, photoactivated chemical reactions, and quantum materials.

\begin{suppinfo}
\end{suppinfo}

\section{Acknowledgements}

This work was supported by the NSF CAREER award through Grant No.~DMR-1945098 (V.V.). In part, this work was supported by the NSF Quantum Foundry through QAMASE-i program Award No.~DMR-1906325 (V.V). This work was partially supported by a \emph{Obra Social ``La Caixa''} graduate fellowship (ID 100010434), with code LCF/BQ/AA16/11580047 (C.M.Z.). N.M.T. and S.C. are grateful for support from NASA Ames Research Center and support from the AFRL Information Directorate under Grant No.~F4HBKC4162G001.
The calculations were performed as part of the XSEDE computational Project No. TG-CHE180051 and TG-MCA93S030. Use was made of computational facilities purchased with funds from the National Science Foundation (CNS-1725797) and administered by the Center for Scientific Computing (CSC). The CSC is supported by the California NanoSystems Institute and the Materials Research Science and Engineering Center (MRSEC; NSF DMR 1720256) at UC Santa Barbara. We thank Mark Babin, Blake Erickson and Diptarka Hait for helpful discussions. 

\providecommand{\latin}[1]{#1}
\makeatletter
\providecommand{\doi}
  {\begingroup\let\do\@makeother\dospecials
  \catcode`\{=1 \catcode`\}=2 \doi@aux}
\providecommand{\doi@aux}[1]{\endgroup\texttt{#1}}
\makeatother
\providecommand*\mcitethebibliography{\thebibliography}
\csname @ifundefined\endcsname{endmcitethebibliography}
  {\let\endmcitethebibliography\endthebibliography}{}

\end{document}